\documentclass[aip, apl, reprint]{revtex4-1}

\usepackage{outlines}
\usepackage{siunitx}
\usepackage{graphicx}
\usepackage{amsmath}
\usepackage{blindtext}
\usepackage{sidecap}
\usepackage{lineno}
\usepackage{epstopdf}
\usepackage[T1]{fontenc}
\usepackage{hyperref}
\usepackage{natbib}
\usepackage{sidecap}
\hypersetup{
  colorlinks   = true,
  urlcolor     = blue, 
  linkcolor    = red, 
  citecolor   = blue
}

\begin{document}

\title{Automated tuning of inter-dot tunnel couplings in quantum dot arrays}

\author{C.~J.~van Diepen}
\affiliation{QuTech, 
Delft University of Technology, P.O. Box 5046, 2600 GA Delft, The Netherlands}
\affiliation{Kavli Institute of Nanoscience, 
Delft University of Technology, P.O. Box 5046, 2600 GA Delft,  The Netherlands}
\affiliation{Netherlands Organisation for Applied Scientific Research (TNO), P.O. Box 155, 2600 AD Delft, The Netherlands}
\author{P.~T.~Eendebak}
\affiliation{QuTech, 
Delft University of Technology, P.O. Box 5046, 2600 GA Delft, The Netherlands}
\affiliation{Netherlands Organisation for Applied Scientific Research (TNO), P.O. Box 155, 2600 AD Delft, The Netherlands}
\author{B.~T.~Buijtendorp}
\affiliation{QuTech, 
Delft University of Technology, P.O. Box 5046, 2600 GA Delft, The Netherlands}
\affiliation{Kavli Institute of Nanoscience, 
Delft University of Technology, P.O. Box 5046, 2600 GA Delft, The Netherlands}
\affiliation{Netherlands Organisation for Applied Scientific Research (TNO), P.O. Box 155, 2600 AD Delft, The Netherlands}
\author{U.~Mukhopadhyay}
\author{T.~Fujita}
\affiliation{QuTech, 
Delft University of Technology, P.O. Box 5046, 2600 GA Delft, The Netherlands}
\affiliation{Kavli Institute of Nanoscience, 
Delft University of Technology, P.O. Box 5046, 2600 GA Delft, The Netherlands}
\author{C.~Reichl}
\author{W.~Wegscheider}
\affiliation{Solid State Physics Laboratory, ETH Z\"urich, 8093 Z\"urich, Switzerland}
\author{L.~M.~K.~Vandersypen}
\affiliation{QuTech, 
Delft University of Technology, P.O. Box 5046, 2600 GA Delft, The Netherlands}
\affiliation{Kavli Institute of Nanoscience, 
Delft University of Technology, P.O. Box 5046, 2600 GA Delft, The Netherlands}
\date{\today}

\begin{abstract}
Semiconductor quantum dot arrays defined electrostatically in a 2D electron gas provide a scalable platform for quantum information processing and quantum simulations. For the operation of quantum dot arrays, appropriate voltages need to be applied to the gate electrodes that define the quantum dot potential landscape. Tuning the gate voltages has proven to be a time-consuming task, because of initial electrostatic disorder and capacitive cross-talk effects. Here, we report on the automated tuning of the inter-dot tunnel coupling in a linear array of gate-defined semiconductor quantum dots. The automation of the tuning of the inter-dot tunnel coupling is the next step forward in scalable and efficient control of larger quantum dot arrays. This work greatly reduces the effort of tuning semiconductor quantum dots for quantum information processing and quantum simulation.
\end{abstract}

\maketitle

Electrostatically defined semiconductor quantum dots are actively studied as a platform for quantum computation \cite{Loss1998, Hanson2007, Zwanenburg2013} and quantum simulation. \cite{Byrnes2008, Barthelemy2013} Control over the inter-dot tunnel coupling is a key ingredient for both applications. Via control over the tunnel coupling we have control over the exchange coupling, which is vital for realizing the various proposals for spin-based qubits. \cite{Loss1998, Levy2002, DiVincenzo2000a} Based on the natural description of semiconductor quantum dots in terms of the Fermi-Hubbard model, control over the tunnel coupling allows for analog simulations to explore the physics of interacting electrons on a lattice. \cite{Yang2011, Hensgens2017} \\
\indent An obstacle for the efficient use of semiconductor quantum dots are the background charged impurities and variations in the gate patterns, which lead to a disordered potential landscape. Initial disorder can be compensated for by applying individually adjusted gate voltages. Additionally, even though gates are designed to specifically control a chemical potential or a tunnel coupling, in practice capacitive coupling induces cross-talk from all gates to dot chemical potentials and tunnel couplings. The disorder and cross-talk increase the complexity of tuning up ever larger dot arrays. The effort of tuning can be reduced by automation based on image processing. Earlier work on automation of tuning for semiconductor quantum dots has shown that it is possible to automatically form double quantum dots with a sensing dot, and to find the single electron regime in the double dot. \cite{Baart2016b} More recently, these automated tuning routines were used to determine the initialization, read-out and manipulation points for a singlet-triplet qubit. \cite{Botzem2018} Automated control over the inter-dot tunnel coupling is an important next step forward in control for scaling up the number of spin qubits in semiconductor quantum dots. \\
\indent In this Letter, we present and implement a computer-automated algorithm for the tuning of the inter-dot tunnel coupling in semiconductor quantum dot arrays. The algorithm consists of two parts. Part I determines a virtual barrier gate, which corresponds to a linear combination of voltages to apply on multiple gates in order to adjust the tunnel barrier without influencing the chemical potentials in the dots. To determine such a virtual barrier gate we model and fit the capacitive anti-crossings measured in charge stability diagrams. Part II tunes the tunnel coupling using a feed-back loop, which consists of stepping the virtual barrier gate value and measuring the tunnel coupling, until the tunnel coupling converges to the target value. To measure the tunnel coupling we use two methods. The first method is based on photon-assisted tunneling\cite{Oosterkamp1998} (PAT), while the second method is based on the broadening of the inter-dot transition line. \cite{DiCarlo2004} We describe the algorithm and demonstrate its power by automatically tuning the tunnel coupling to a target value for two double dots. We show results for tuning both to higher and lower tunnel couplings for several different initial values, both for a single electron and for two electrons on the double dot. \\
\begin{figure*}
\includegraphics[width=.95\textwidth]{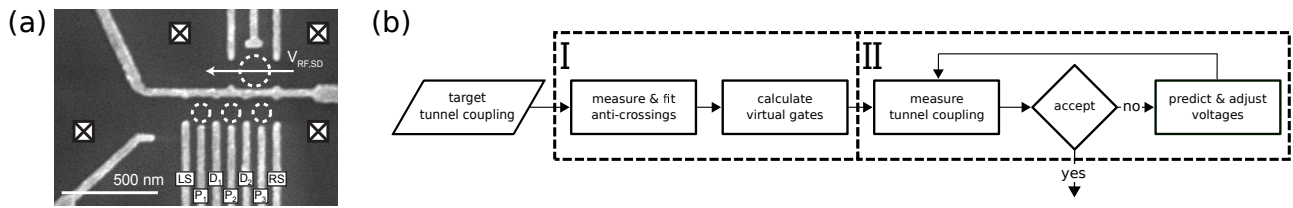}
\caption{(a) A scanning electron microscopy image of a device nominally identical to the one used for the measurements. The three smaller dashed circles indicate the positions of the dots in the array. The larger dashed circle indicates the location of the sensing dot. Squares indicate Fermi reservoirs, which are connected to ohmic contacts. (b) A flowchart of the automated tunnel coupling tuning algorithm. The dashed boxes indicate the two parts of the algorithm.\label{fig:algo_flow}}
\end{figure*}
\indent The platform used for the demonstration of the algorithm is a linear triple quantum dot device. \cite{Baart2015} A scanning electron microscopy image of a device similar to the one used in our experiment is shown in Fig.~\ref{fig:algo_flow}(a). By applying voltages on gate electrodes on the surface of a GaAs/AlGaAs heterostructure, we shape the potential landscape in the two-dimensional electron gas 85 nm below the surface. Gates $LS$ and $RS$ are designed to control the tunnel couplings to the left and right reservoir, respectively. Additionally, plunger gates, $P_i$, are designed to control the chemical potential of dot $i$, and barrier gates, $D_i$, are designed to control the inter-dot tunnel coupling between dot $i$ and dot $i+1$. The device allows for the formation of three quantum dots in a linear configuration, which are indicated with three white dashed circles in the bottom part of Fig.~\ref{fig:algo_flow}(a) and one additional dot, indicated with the larger white dashed circle in the upper part. We refer to this additional dot as the sensing dot (SD), because it is operated as a charge sensor, utilizing its capacitive coupling to the three other quantum dots. One of the SD contacts is connected via a bias-tee to a resonator circuit, permitting fast read-out of the charge configuration in the bottom dots, by measuring the SD conductance with radio-frequency reflectometry. To optimize the sensitivity of the charge sensor, we operate the SD half-way on the flank of a Coulomb peak. Automation on the tuning of the sensing dot for read-out was already shown in Ref. \citenum{Baart2016b}. One of the bottom gates, $P_2$, is connected to a microwave source, used for PAT measurements. \\
\indent As starting point for our algorithm, we assume that the device is tuned near an inter-dot charge transition. Such a starting point can be obtained from a computer-automated tuning algorithm \cite{Baart2016b}. We also require a rough estimate of the electron temperature for the modelling of charge transition line widths. For the PAT measurements, we calibrated the microwave power such that we only observe single-photon lines. \cite{Oosterkamp1998} \\
\indent Part I of the algorithm, see Fig.~\ref{fig:algo_flow}(b), determines the virtual plunger and barrier gates by measuring the cross-capacitance matrix (see supplementary material II), which describes the capacitive couplings from gates to dot chemical potentials. To determine this matrix we measure charge stability diagrams with charge sensing and fit the avoided crossing with a classical model (supplementary Fig.~1). The fitting of the anti-crossings is based on finding the minimum of the sum over all pixels of the difference between the processed data and a two-dimensional classical model of the avoided crossing (see supplementary material III). From the fit of the anti-crossing, we obtain the slopes of all five transition lines: four addition lines, where an electron moves between a reservoir and a dot, and the inter-dot transition line, where a charge moves from one dot to the other. We fit the anti-crossing to charge stability diagrams measured for any combination of $P_i$, $P_{i+1}$ and $D_i$ over a range of $\SI{40}{\milli\volt}$ around the starting point, to fill in the entries of the cross-capacitance matrix. From the inverse of this matrix we obtain both the virtual barrier, $\widetilde{D}_i$, and the virtual plungers, $\widetilde{P}_i$ and $\widetilde{P}_{i+1}$. The effectiveness of this basis transformation in voltage-space becomes clear from the right angles between addition lines in the charge stability diagram in the 2D-scan of $\widetilde{P}_i$ and $\widetilde{P}_{i+1}$ in Fig.~\ref{fig:models}(a). The anti-crossing fit also provides the voltages at the center position on the inter-dot transition line, indicated with the white dot. The white dotted line indicates the detuning axis, which will be used as a scanning axis in the second half of the algorithm. \\
\begin{SCfigure*}
\includegraphics[width=0.47\textwidth]{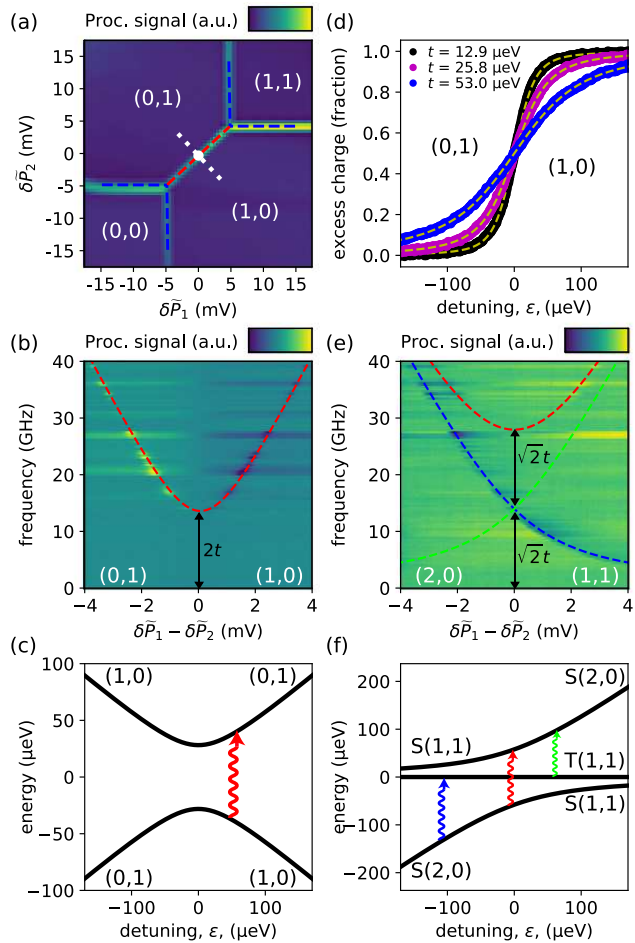}
\caption{In all subfigures, $(N_1, N_2)$ indicates charge occupation of the left and middle dot, with no dot formed on the right. (a) A double quantum dot charge stability diagram, showing the processed sensing dot signal as a function of virtual plunger gate voltages. The fitted anti-crossing model is indicated with dashed lines. The detuning axis is indicated with the white dotted line and the center point on the inter-dot transition line with a white dot. (b) Photon-assisted tunneling measurement showing the charge detector signal (background subtracted) as a function of frequency and inter-dot detuning at the (0,1) to (1,0) transition. The red dashed line is a fit of the form $hf=\sqrt{\varepsilon^2+4t^2}$. The detuning lever arm is extracted from the slope of the hyperbola in the large detuning limit. (c) The energy level diagram for one-electron occupation. The eigenenergies are $\pm \frac{1}{2} \sqrt{\varepsilon^2+4t^2}$. A microwave photon (red wiggly arrow) can induce a transition (and potentially tunnelling between the dots) when the difference between the energy levels corresponds to the photon energy (PAT). (d) Excess charge extracted from a fit to the sensing dot signal as a function of $\varepsilon$ for different $t$, measured by scanning over the detuning axis for the single-electron occupation. The model used to fit to the SD signal is $V(\varepsilon) = V_0 + \delta V Q(\varepsilon) + \left[ \frac{\delta V}{\delta \varepsilon} \big|_{Q=0} + \left( \frac{\delta V}{\delta \varepsilon} \big|_{Q=1} - \frac{\delta V}{\delta \varepsilon} \big|_{Q=0} \right) Q(\varepsilon) \right] \varepsilon$. Here $V_0$ is the background signal, $\delta V$ is a measure of the charge sensitivity, $Q$ the excess charge as a fraction of the electron charge and $\frac{\delta V}{\delta \varepsilon}$ the gate-sensor coupling when $\varepsilon$ is varied. \cite{Hensgens2017} (e) Photon-assisted tunneling measurement similar to (b) but for the inter-dot transition from (2,0) to (1,1). Coloured dashed lines are fits to the measured data. (f) The energy level diagram for the two electron transition. Coloured wiggly arrows indicate microwave photon excitations. The energy levels are given by $\frac{\varepsilon}{2} \pm \frac{1}{2} \sqrt{\varepsilon^2 + 8 t^2}$ for the singlets and are $0$ for the degenerate triplets. \label{fig:models}}
\end{SCfigure*}
\indent Before describing part II of the algorithm let us first explain the two methods we use to measure the tunnel coupling. The first method is PAT, see Fig.~\ref{fig:models}(b) and (e), which is based on the re-population of states induced by a microwave field. We can observe the re-population using the sensing dot, when the different states correspond to different charge configurations. While varying the frequency of the microwave source, we observe resonance peaks when the frequency is equal to the energy difference between two states. By scanning over the detuning axis and finding the resonance peaks we perform microwave spectroscopy to map out (part) of the energy level diagram, from which we determine the tunnel coupling. We obtain the tunnel coupling by using a fitting procedure that consists of three steps. First we process the data per microwave frequency, mainly subtracting a smoothed background signal taken when the microwave source is off. Second we find the extrema in this processed signal per microwave frequency and last we fit the curve(s) that connects the extrema using a model of the energy level diagram. For the PAT measurement with a single electron as shown in Fig.~\ref{fig:models}(b), we model the system in terms of two levels with energies as shown in Fig.~\ref{fig:models}(c). The resonance curve is then described by $hf = \sqrt{\varepsilon^2 + 4t^2}$, where $h$ is Planck's constant, $f$ the applied microwave frequency, $t$ the inter-dot tunnel coupling and $\varepsilon$ the detuning, which is given by  $\alpha (\delta \widetilde{P}_i - \delta \widetilde{P}_{i+1})$, with $\alpha$ the lever arm, a conversion factor between voltage and energy scales.\cite{Oosterkamp1998} If two electrons occupy the two dots at zero magnetic field, there are three relevant energy levels, two corresponding to singlet states and the other to threefold degenerate triplet states, see Fig.~\ref{fig:models}(f). This level structure results in three possible transitions, \cite{Hanson2007} with energies described by $hf = \frac{\varepsilon}{2} \pm \frac{1}{2} \sqrt{\varepsilon^2 + 8 t^2}$ and $hf = \sqrt{\varepsilon^2 + 8t^2}$. In the measurement shown in Fig.~\ref{fig:models}(e) we only observe two out of the three transitions. This we explain by observing that the thermal occupation of the lowest excited state is negligible. We note that some PAT transitions involve a spin-flip, which is mediated by spin-orbit interaction and a difference in the Overhauser fields between the two dots. \cite{Schreiber2011} The variation in intensity for different horizontal lines in Fig.~\ref{fig:models}(b) and (e) is caused by the frequency dependence of the transmission of the high-frequency wiring. One could compensate for this by adjusting the output power of the microwave source per frequency. The blue tails in Fig.~\ref{fig:models}(e) are caused by sweeping gate voltages at a rate which is of the same order of magnitude as the triplet-singlet relaxation rate. This was confirmed by inverting the sweep direction and observing that the blue tails appear on the other side of the transition line. \\
\indent The second method to measure the tunnel coupling is based on the broadening of the inter-dot transition line \cite{DiCarlo2004}, see Fig.~\ref{fig:models}(d). The broadening reflects a smoothly varying charge distribution when scanning along the detuning axis, caused by the tunnel coupling via the hybridisation of the relevant states and the temperature through the thermal occupation of excited states. For the single-electron case, the average excess charge on the left (right) dot is given by
\begin{equation}
\label{eqn:exc_ch}
Q = \frac{1}{\mathcal{Z}} \sum_n (c_n \mathrm{e}^{-E_n/k_B T_e}),
\end{equation}
with $\mathcal{Z}$ the partition function, $c_n = \frac{1}{2} \mp \varepsilon / E_n$ the probability of finding the excess charge on the left (right) dot for the eigenstate with energy $E_n$ and the thermal energy $k_B T_e \approx \SI{10.5}{\micro eV}$, with $T_e$ the effective electron temperature. An analogous expression applies to the two-electron case, with $c_n = 0$ for the triplets and $c_n = \frac{1}{2}(1 \pm \varepsilon / \sqrt{\varepsilon^2+8t^2})$ for the hybridized singlets. The lever arm used for measuring the tunnel coupling from the broadening of the inter-dot transition line is obtained from PAT, but could also be measured with Coulomb diamonds or bias triangles. \cite{Hanson2007} Based on Eqn.~\ref{eqn:exc_ch} we obtain the model for the charge sensor response when scanning over the detuning axis, see the caption of Fig.~\ref{fig:models}. \cite{Hensgens2017} \\
\begin{SCfigure*}
\includegraphics[width=.7\textwidth]{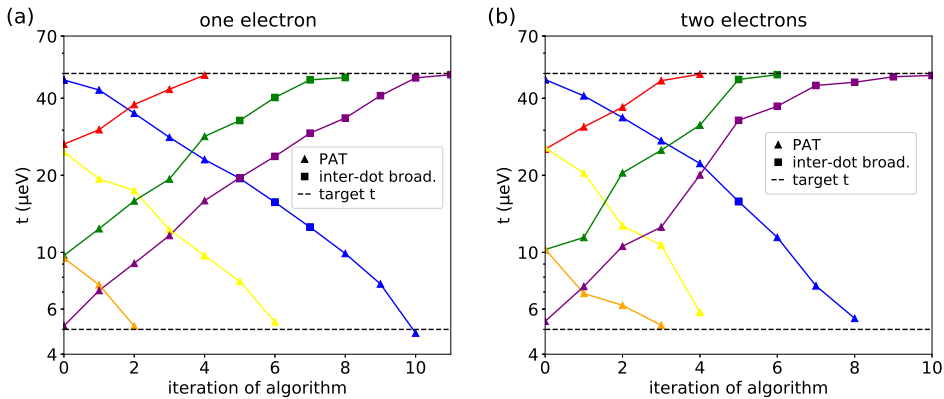}
\caption{Results of the algorithm for a transition involving (a) single-electron states and (b) two-electron states. Each panel shows results for different runs of the algorithm indicated by different colors. Triangles indicate tunnel coupling values measured with PAT and squares indicate measurements based on inter-dot transition line broadening. Solid lines are added as a guide to the eye. The black dashed lines indicate the high and low target tunnel coupling values. \label{fig:algo}}
\end{SCfigure*}
\indent Here we compare the two methods for extracting the tunnel coupling. An advantage of the method based on the broadening of the inter-dot transition line is that it is about two orders of magnitude faster than PAT (see Table I in the supplementary material), because it is effectively a single scan over the detuning axis while PAT is a series of scans over the detuning axis for different microwave frequencies. Another difference is in the range of tunnel couplings over which the two methods work well. For PAT the upper limit depends on the maximum frequency that the microwave source can produce. We expect that the lower limit for PAT is determined by charge noise, resulting in broadening of the PAT peaks. With PAT, we were able to automatically measure tunnel coupling values as low as $\SI{5}{\micro \electronvolt}$. The lower limit for the inter-dot transition broadening method is set by the effective electron temperature, $k_B T_e$, here $\approx \SI{10.5}{\micro eV}$. \cite{DiCarlo2004} The upper limit for this method is that for very large tunnel couplings, the broadening of the inter-dot transition line extends to the boundaries of the charge stability region. In the measurements shown here, we did not come close to this upper limit, but tunnel couplings up to $\SI{75}{\giga\hertz} \approx \SI{300}{\micro\electronvolt}$ have been measured with the inter-dot transition line broadening method.\cite{Hensgens2017}. We observe that the two methods are in good correspondence with one another, i.e. the difference between the two is smaller than 10\% of their average value (see supplementary material V). Measurement errors are usually smaller than the accuracy in target tunnel coupling we are interested in, while potential outliers will typically be caused by unpredictable charge jumps. \\
\indent Now, let us describe part II of the algorithm, see Fig.~\ref{fig:algo_flow}(b), which performs a feedback loop. For each iteration the virtual barrier gate value is adjusted and the tunnel coupling is measured. Before the first step of the algorithm we measure the tunnel coupling with PAT. If we are not yet within $\SI{1}{\micro eV}$, of the target tunnel coupling value, we step the virtual barrier gate value with step size equal to the maximal step size in the positive direction if the tunnel coupling is too low and vice versa. We limit the barrier gate step size to $\SI{20}{\milli\volt}$ such that the position of the anti-crossing can again be located automatically by fitting the anti-crossing model. For larger step sizes, the position of the anti-crossing becomes harder to predict due to non-linearities. After stepping the virtual barrier gate we measure the tunnel coupling again using PAT. Then we have measured the tunnel coupling for two settings and we determine the next step for the virtual barrier, by predicting the voltage required to reach the target value from an exponential fit  \cite{Simmons2009, Hensgens2017} to the measured tunnel couplings and their respective virtual barrier values (we thereby force the exponential to go to zero for very negative barrier voltages). After the tunnel coupling has been measured five times with PAT we also have five measured lever arm values for different gate voltages. The small differences in lever arm we interpret as caused by small shifts in the dot positions with the gate voltages. We predict the lever arm for other voltages using a linear approximation (see supplementary material V). Using this knowledge of the lever arm, the algorithm can be sped up for the subsequent iterations by measuring the tunnel coupling from inter-dot transition broadening. \\
\indent Following the procedure described above the algorithm automatically tunes the inter-dot tunnel coupling to a target value, within the range of the measurable tunnel coupling values and the achievable values with our gate design and electron occupations. Fig.~\ref{fig:algo} shows the results of the tuning algorithm for various initial and target tunnel coupling values, indicated with different colours. The target tunnel coupling value are indicated with black dashed lines. We clearly see that the algorithm finds the gate voltages that bring the tunnel coupling to the target value, stepwise moving closer. In Fig.~\ref{fig:algo}(a) results for the left pair of dots with a single electron are shown, while Fig.~\ref{fig:algo}(b) shows results for an occupation with two electrons. We have obtained similar results for the second pair of neighbouring dots in the triple dot (see supplementary material VII). The duration of a run of the algorithm mainly depends on the difference between the initial and the final tunnel coupling value, because we limit the maximum step size. The duration typically is in the order of 10 min (see supplementary material VI for more details). \\
\indent In conclusion, we have shown automation of the tuning of the tunnel coupling between adjacent semiconductor quantum dots. Key for this automation were image processing methods to automatically fit the shape of an anti-crossing and to find the shape of the resonance curve in a PAT measurement. The present methods for measuring inter-dot tunnel couplings and the feedback routine can be extended to larger quantum dot arrays. This work demonstrates further automated control over semiconductor quantum dots and is the next step forward in automated tuning of larger quantum dot arrays, necessary for scaling up the number of spin-based qubits implemented with semiconductor quantum dots. \\
\indent The authors acknowledge useful discussions with T.~Hensgens, J.~P.~Dehollain and other members of the Vandersypen group, experimental assistance by C.~A.~ Volk and A.~M.~J.~Zwerver, and technical support by M.~Ammerlaan, J.~Haanstra, S.~Visser and R.~Roeleveld. This work was supported by the Netherlands Organization for Scientific Research (NWO Vici), and the Dutch Ministry of Economic Affairs through the allowance for Top Consortia for Knowledge and Innovation (TKI) and the Swiss National Science Foundation. 

\bibliography{bib_main}

\clearpage

\begin{center}
\textbf{Supplementary Material}
\end{center}

\section{Software and algorithms}
The software was developed using Python~\cite{Python}, SciPy~\cite{SciPy} and the QCoDeS~\cite{QCoDeS} framework. The image processing is performed in pixel coordinates. The parameters of algorithms are given in physical units such as mV. The corresponding parameter in pixel units is then determined by translating the value using the scan parameters. By specifying the parameters in physical units the algorithms remain valid also if scans are made with a different resolution. Of course making scans with a different resolution can lead to differences in rounding of numbers leading to slightly different results.

\section{Virtual gates}
Due to the capacitive coupling from gates to dot chemical potentials and tunnel barriers, changing the voltage applied on one of the gates influences not only one but all of the chemical potentials and tunnel barriers in the potential landscape. To compensate for the cross-talk from gates to chemical potentials we make use of a cross-capacitance matrix. The entries of this matrix correspond to the coupling strengths. The columns in the inverse of the cross-capacitance matrix contain the coefficients for the gate combinations defining the virtual gates. The virtual plungers, $\tilde{P_i}$, which are linear combinations of plungers, $P_i$, control the chemical potential in one dot while leaving the other chemical potentials unaffected. The virtual barrier, $\tilde{D_i}$, changes the inter-dot tunnel coupling without affecting the chemical potentials, hence contains compensation for the effect of the barrier, $D_1$, on the dot chemical potentials. An example of a measured cross-capacitance matrix is
\begin{equation}
\label{eqn: ccmatrix}
\begin{pmatrix}
\delta \widetilde{P}_1 \\ \delta \widetilde{P}_2 \\ \delta \widetilde{D}_1
\end{pmatrix}
= 
\begin{pmatrix}
1.00 & 0.49 & 1.23 \\
0.55 & 0.88 & 1.49 \\
0.00 & 0.00 & 1.00
\end{pmatrix}
\begin{pmatrix}
\delta P_1 \\ \delta P_2 \\ \delta D_1
\end{pmatrix}.
\end{equation}
The upper two rows are scaled such that the top-left entry is one. The left two entries of the bottom row describe the effect of the plunger gates on the tunnel barrier. These entries are set to zero because the PAT and inter-dot line broadening measurements are performed near an inter-dot transition, hence using these methods we could not independently measure the effect of plungers on the tunnel barrier. The last entry of the row for the couplings to the barrier is set to one as we chose the effect of the physical barrier on the virtual barrier to be one-to-one. 

\begin{figure*}[ht]
\includegraphics[width=\textwidth]{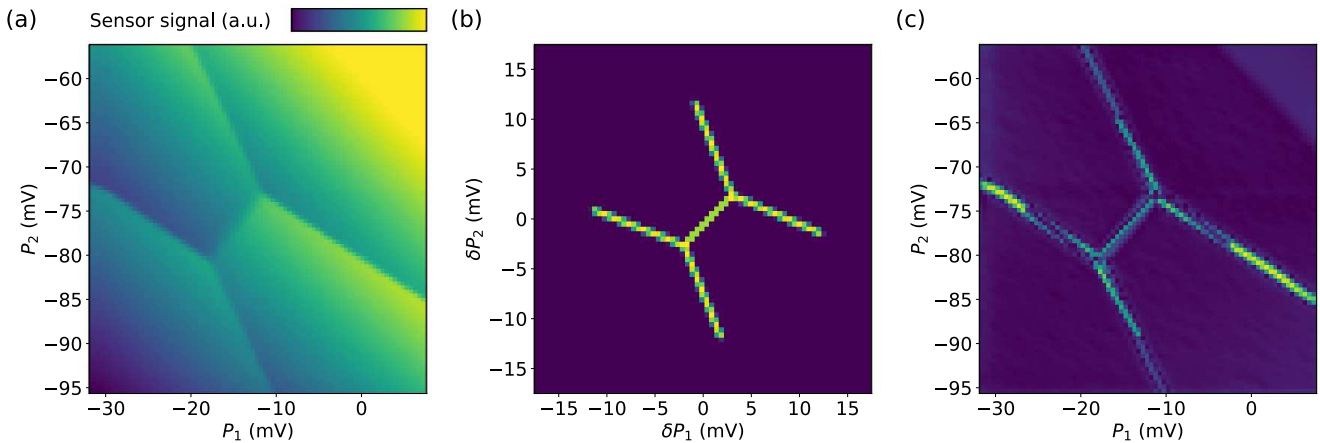}
\caption{(a) A charge stability diagram with an avoided crossing, showing the unprocessed sensor signal. (b) The two-dimensional patch generated based on the classical model of the anti-crossing. (c) Processed sensor signal recorded during the charge stability measurement, from which the model fitted on the anti-crossing is subtracted.  \label{fig:ac_fitting}}
\end{figure*}

\section{Avoided crossing model}
Here, we describe the fitting routine of an avoided crossing in a charge stability diagram, see Fig.~\ref{fig:ac_fitting}(a). For this fitting we developed a two-dimensional, classical model of an avoided crossing which will be explained below. First we describe the processing of the measured data. \\
\indent The first step in the processing is calculating a derivative of the image. This is done by applying a first order Gaussian filter. From the convoluted data we subtract a background signal. This background signal is a third order polynomial fit to the data, which was convoluted with the Gaussian filter. This background subtraction is done to remove the shape of the sensing dot Coulomb peak in the data. Next step is to straighten the measured data into a square image. This straightening ensures that horizontal and vertical directions are equally weighted in the fitting with the 2D model. Then we normalize the signal with its $99$th percentile. 
\begin{equation}
\mathrm{proc. \hspace{2pt} data} = \frac{\left(\mathrm{conv.\hspace{2pt} data}_{\mathrm{pix.}} - \mathrm{bg_{fit}} \right)}{p_{99} (\mathrm{conv.\hspace{2pt} data}_{\mathrm{pix.}} - \mathrm{bg_{fit}})} .
\end{equation}
\indent We developed a classical model of an anti-crossing as observed in charge stability diagrams. The model consists of a two-dimensional patch, see Fig.~S\ref{fig:ac_fitting}(b). The line shapes in this model are based on a truncated cosine. The model has eight parameters, that need to be fit. Two parameters describe the center of the avoided-crossing, as indicated with a white dot in Fig.~\ref{fig:models}(a) in the main text, five parameters describe the angles of the four addition lines and the inter-dot transition line, and one parameter corresponds to the length of the inter-dot transition line. Additional to these eight parameters, which are to be fitted, there are two more parameters, which we fix before the fitting. The first is the typical width of an addition line, which is based on the effective electron temperature. The second parameter is the length of the four line pieces which we fit on the addition lines. These are chosen such that they are significantly larger than the effective electron temperature and smaller than the addition energy. 

The anti-crossing is fit by minimizing the following cost function
\begin{equation}
\mathrm{cost} = \sum_{\mathrm{pixels}} \left[ |\mathrm{proc.\hspace{2pt} data}| - \mathrm{model} \right] ,
\end{equation}
which is the sum over all pixel intensities of the processed data minus the 2D patch of the model. This fitting procedure results in a fit as shown in Fig.~\ref{fig:ac_fitting}(c). 

\begin{figure*}[h]
\includegraphics[width=\textwidth]{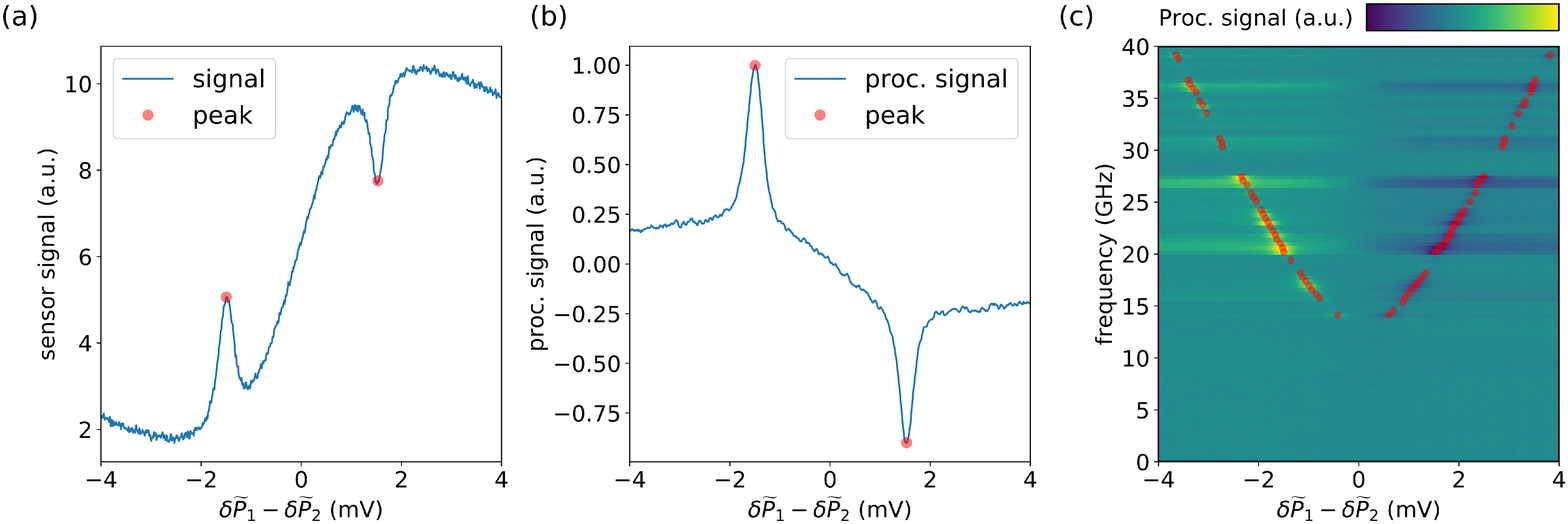}
\caption{(a) Line cut of the sensing dot signal as a function of detuning with the microwave frequency, $f$, at $\SI{20}{\giga\hertz}.$ (b) At the same microwave frequency, but here showing the processed sensing dot signal, hence with the smoothed background subtracted and with the remaining signal normalized. (c) A PAT measurement, showing the processed sensing dot signal, with the red dots indicating the detected peaks. \label{fig:supp_pat}}
\end{figure*}

\section{Photon assisted tunnelling fitting}
Here we explain the fitting procedure for the photon assisted tunnelling measurements. The PAT fitting procedure consists of three steps: processing the data, detecting the detuning values of the resonance peaks per microwave frequency, see Fig.~\ref{fig:supp_pat}(a), and fitting the curve describing the energy difference to the detected peaks. 

The processing of the data is done per horizontal line in a PAT measurement, i.e.\ per applied microwave frequency. The first step is the subtraction of a background signal. We measure the background signal with a scan over the detuning axis while the microwave source is off, note that this is the same scan we would do when we want to measure the tunnel coupling based on the broadening of the inter-dot transition line. Before subtracting the background signal we smoothen both the signal and the background signal with a Gaussian filter with $\sigma$ set to five pixels. After the background subtraction we subtract the average of the signal and rescale it, resulting in the data as shown in Fig.~\ref{fig:supp_pat}(b). 

We detect the resonance peaks as extrema in the processed signal. Just as for the processing of the data, the peak detection is done per horizontal line. First we find the maximum and minimum per horizontal line. We heuristically determined a threshold for the detected extrema based on the difference in signal for the two charge configurations and the noise level. We filter the extrema by only accepting the detected peaks which have an absolute value higher then the threshold, note here that we already normalized the processed signal. 

To the filtered extrema, indicated with red dots in Fig.~\ref{fig:supp_pat}(c), we fit the energy transition model, which is described in the main text. This fitting procedure results in fits as shown in Fig.~\ref{fig:models}(b) and (e) in the main text.

\section{Compare PAT and inter-dot transition line broadening}
\begin{SCfigure*}
\includegraphics[width=.67\textwidth]{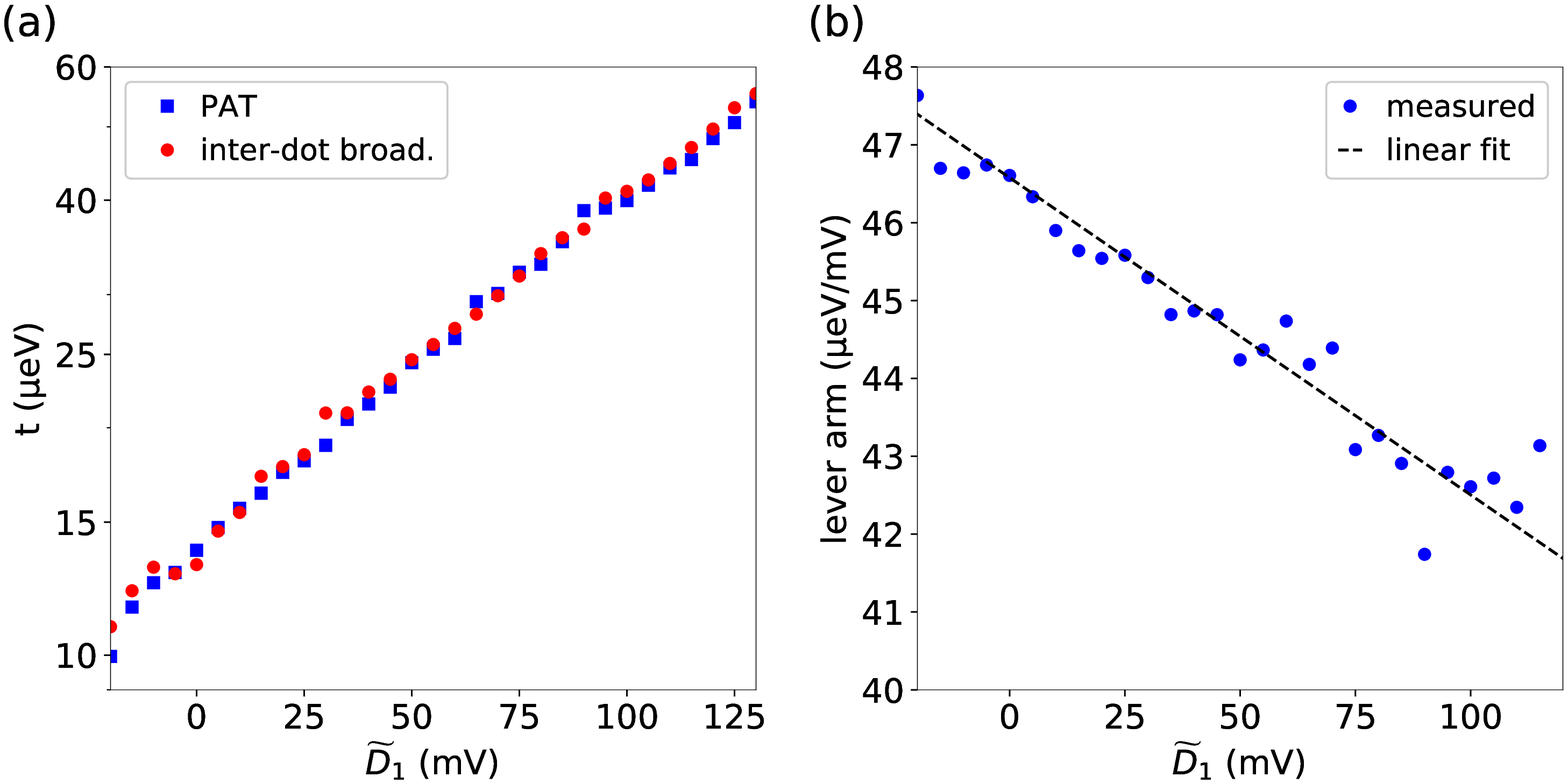}
\caption{(a) The tunnel coupling plotted against the virtual barrier gate voltage. Blue triangles indicate measurements with PAT while filled red circles indicate measurements based on the broadening of the inter-dot transition line. The effective electron temperature used for the inter-dot line fitting is $\SI{10.5}{\micro eV}$. (b) The lever arm measured with PAT plotted against the virtual barrier gate voltage. We fitted a linear relation to the observed trend, which indicates a slight decrease in lever arm as we increase the virtual barrier voltage. \label{fig:charac}}
\end{SCfigure*}

We compare the tunnel coupling measurements based on PAT and those based on the inter-dot transition line broadening to check that they are in agreement with one another. We use both methods to measure the tunnel coupling over a range of tunnel coupling values for which both methods are reliable. These measurements were done in the single electron occupation regime. In Fig.~\ref{fig:charac} we show measured tunnel couplings by both the PAT method and the method based on broadening of the inter-dot line. The lever arm we used for the inter-dot line broadening measurement is taken from the PAT measurement at that virtual barrier gate voltage, see Fig.~\ref{fig:charac}(b). 

\section{Time required to run the algorithm}
In Table~\ref{tab:timings} an overview is given with the approximate times used for the different routines of the tuning algorithm and the tuning algorithm itself. The measurement time for the anti-crossing measurement is lower if the gates on which the voltages are swept are connected to high-frequent lines, hence can be swept with an a.c. signal, or relatively longer if the voltages can only be changed by stepping a d.c. voltage. For the triple dot device used for the demonstration of the algorithm only the plunger gates were connected to high-frequent lines.
\begin{table}
\begin{tabular}{c|c}
\textbf{Measurement}& \textbf{Time} \\ \hline
Anti-crossing & 5 s / 1 min \\
PAT & 1 min \\
POL & 2 s \\
Tuning alg. & 10 min.
\end{tabular}
\caption{The approximate time used per type of measurement. The fitting time is included in the shown durations. \label{tab:timings}}
\end{table}
\section{Additional algorithm results}
\begin{SCfigure*}
\includegraphics[width=.67\textwidth]{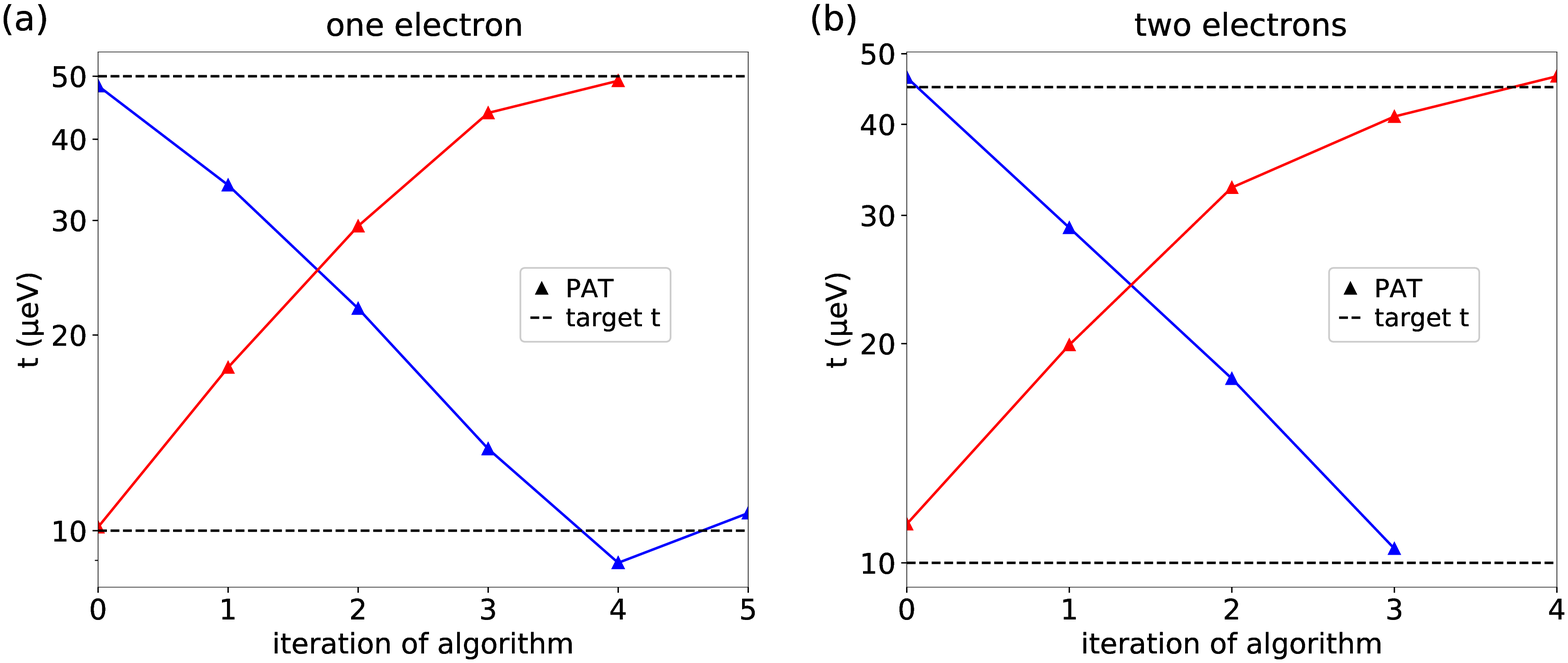}
\caption{Results of the algorithm for a transition involving  single-electron states (a) and (b) two-electron states. These results are for the automated tuning of the tunnel coupling between the pair of dots on the right side of the triple dot device. Each panel shows results for different runs of the algorithm indicated by different colors. Triangles indicate tunnel coupling values measured with PAT. Solid lines are added as a guide to the eye. In this case tune tuning algorithm converged to the target value before switching the inter-dot broadening method. The black dashed lines indicate the high and low target tunnel coupling values. \label{fig:supp_algo_add}}
\end{SCfigure*}

In this section we present additional results of the computer automated tuning algorithm. Fig.~\ref{fig:supp_algo_add} shows the results of the tuning algorithm on the tunnel coupling between the right pair of dots. Again, different colours indicate results of the algorithm for different initial and target tunnel coupling values. Fig.~\ref{fig:supp_algo_add}(a) shows results for the pair of dots with a single electron, while Fig.~\ref{fig:supp_algo_add}(b) shows results for an occupation with two electrons. 

\end{document}